\documentclass[10pt,prb,aps,superscriptaddress,amsmath,amssymb,twocolumn,floatfix,longbibliography]{revtex4-1}
\usepackage{mathtools}
\usepackage{mathrsfs, amsmath, wasysym}
\usepackage{bbold}
\usepackage{dsfont}
\usepackage{braket,mleftright}
\usepackage[mathscr]{eucal}
\usepackage{xfrac}
\usepackage{graphicx}
\usepackage{dcolumn}
\usepackage{bm}
\usepackage{color}
\usepackage{tabularx}
\usepackage{natbib}
\usepackage[bookmarksopen, colorlinks,linkcolor=blue,citecolor=blue,urlcolor=blue]{hyperref}
\usepackage[normalem]{ulem}
\usepackage[all]{hypcap}
\usepackage[dvipsnames,usenames,table]{xcolor}

\newcommand{\bg}{\begin{gather}}
\newcommand{\eg}{\end{gather}}

\newcommand{\be}{\begin{equation} }
\newcommand{\ee}{\end{equation}}
\newcommand{\beq}{\begin{eqnarray}}
\newcommand{\eeq}{\end{eqnarray}}

\newcommand{\bk}{{\bf k}}

\newcommand{\bA}{{\bf A}}

\newcommand{\br}{{\bf r}}
\newcommand{\bR}{{\bf R}}

\newcommand{\Hbdg}{\mathcal{H}_{\text{BdG}}}

\begin{document}

\title{Making $s$-wave superconductors topological  with magnetic field
}

\author{Daniil S. Antonenko}
\email{daniil.antonenko@yale.edu}
\affiliation{Department of Physics, Yale University, New Haven, Connecticut 06520, USA}
\author{Liang Fu}
\affiliation{Department of Physics, MIT, Cambridge, Massachusetts 02139, USA}
\author{Leonid I. Glazman}
\affiliation{Department of Physics, Yale University, New Haven, Connecticut 06520, USA}
\date{\today}

\begin{abstract}
We show that a two-dimensional $s$-wave superconductor may become topological in the presence of a magnetic field that leads to the formation of an Abrikosov vortex lattice.
Below the upper critical field, a  superconducting state with a nontrivial even topological number emerges, which we call the Abrikosov-Chern superconducting state. Deeper in the superconducting domain, the topological number changes in steps, always remaining even and thus not supporting Majorana states, and eventually reaches zero.
Our theory uncovers the nature of evolution from an integer quantum Hall state having a cyclotron gap above the upper critical field to the topologically trivial $s$-wave superconductor carrying finite-energy Caroli--de Gennes--Matricon levels at low field. Topological transitions manifest as changes in the number of edge modes, detectable through tunneling spectroscopy and thermal or spin transport measurements.
\end{abstract}

\maketitle

\section{Introduction}

In the field of topological superconductivity, most of the previous efforts have been focused  \cite{Sato-TopoSCReview} on chiral $p$-wave or $d$-wave superconductivity, either as an intrinsic order  \cite{Ivanov2001} or due to the proximity effect in  specially designed systems  \cite{Oreg, Lutchyn, Beenakker-MajoranaFromSpirals}. 
It was motivated by the predicted occurrence of exotic edge modes, Majorana fermions, and their potential in quantum computing  \cite{Beenakker-Majoranas}. 

On the contrary, the conventional $s$-wave superconductors are perceived to be topologically trivial. The energy spectrum of excitations seen, {\sl e.g.}, in tunneling spectra is characterized by a ``hard'' gap with no states inside it. Applying a relatively weak magnetic field introduces dilute vortices in a superconductor  \cite{Abrikosov_vortex_lattice}. The vortex cores carry Caroli--de Gennes--Matricon (CdGM) levels  \cite{CdGM}. These have low, yet still finite, energy. It is hard to expect that the appearance of vortices induced by a weak magnetic field would lead to topological superconductivity with chiral edge modes, despite the magnetic field breaking the time-reversal symmetry.
However, we may consider the problem from the other limit of high magnetic field and weak superconducting pairing.
The presence of well-defined chiral edge modes in the integer quantum Hall effect (IQHE) is protected by a finite bulk gap, associated with the cyclotron energy $\hbar\omega_c$. Introduction of a weak, compared to $\hbar \omega_c$,  superconducting pairing field $\Delta$ breaks particle number conservation but the chiral edge states (of Bogoliubov quasiparticles) and the quantized thermal Hall effect  \cite{ReadGreen_2000} should persist until a topological gap-closing transition occurs in the bulk. 
The number of edge modes is actually large in the semiclassical limit of a large Fermi energy $E_F / \omega_c \gg 1$. This tension between the two limits, a large number $N \sim E_F/\hbar\omega_c$ of chiral edge modes at the onset of superconductivity versus no edge modes in the case of a fully developed superconducting gap,
sets the stage for the following question: Do edge modes vanish upon increasing the superconducting gap, and if so, how do they vanish? Or equivalently, does the topology of an $s$-wave superconductor change with the magnetic field, and if so, how does it change?

\begin{figure}	\includegraphics[width=0.45\textwidth]{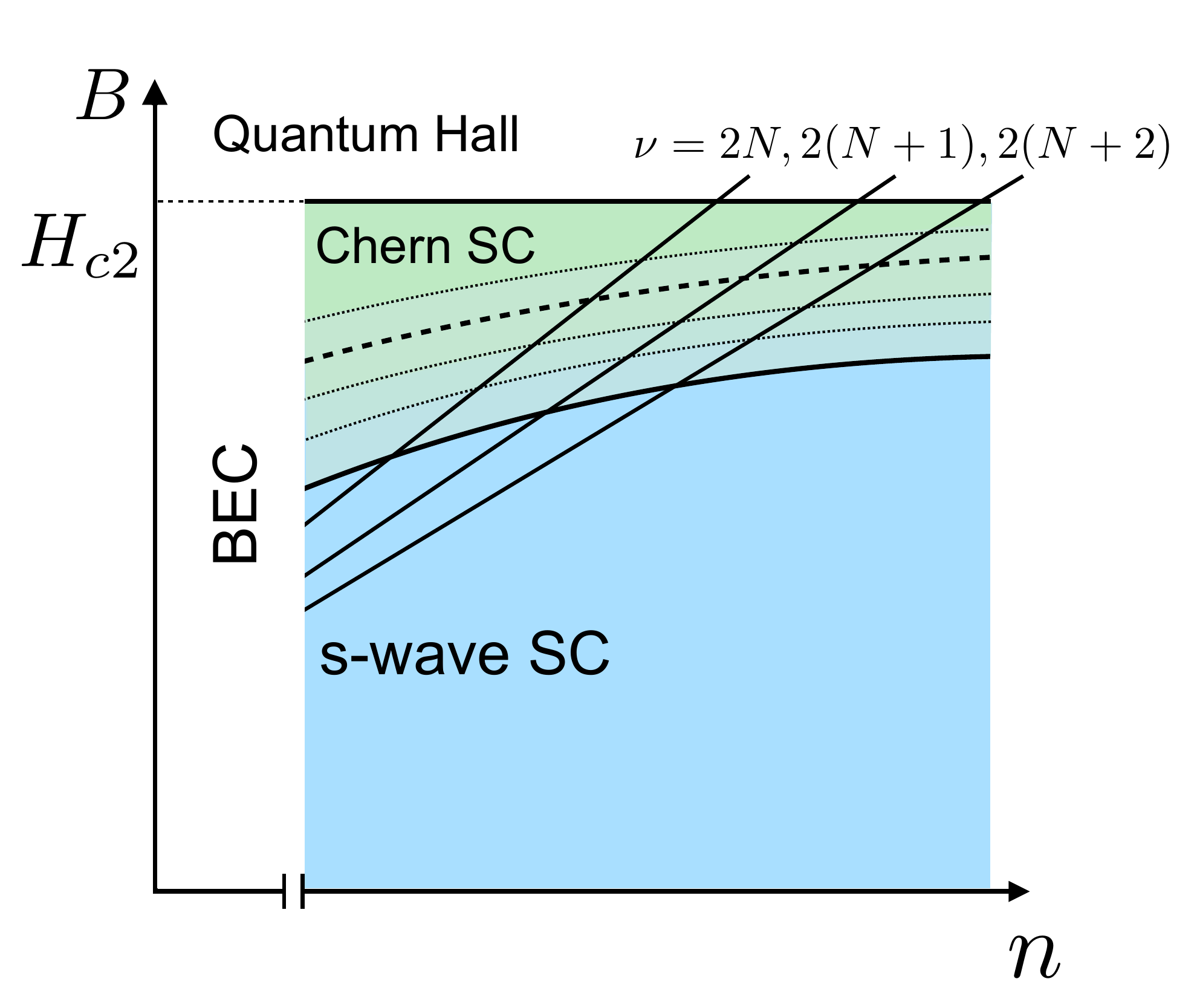}
	\caption{ Phase diagram of a two-dimensional superconductor in an out-of-plane magnetic field $B$. A fixed even value of the filling factor is maintained by a simultaneous variation of $B$ and the charge carrier density $n$ along one of the solid oblique lines. 
    A topologically nontrivial state survives below the  upper critical field $H_{c2}$ {(``Chern SC'')}. The topological number changes in a series of steps, accompanied by the spectral gap closures (dotted lines), eventually bringing the system to a topologically-trivial state, which is shown in blue and delineated by a solid line. The dashed line indicates the boundary of the order parameter fluctuation region. At low density $n$, a crossover to the BEC regime occurs.
 }
\label{fig:phase_diagram}
\end{figure}

Our work provides a detailed answer to these questions. We investigate the evolution of the ground state and low-energy excitations of a two-dimensional electron system upon increasing 
the ratio $\Delta/\hbar\omega_c$ at a fixed, even value of the filling factor $\nu = 2 E_F / (\hbar\omega_c)$, corresponding to filling $\nu/2$ Landau levels for each spin projection. The pairing field may originate from the interactions within the system, as in the experiment with a  gate-controlled twisted trilayer graphene  \cite{ParkNature}. In this case, to vary $\Delta$ at fixed $\nu$, one needs to simultaneously decrease the charge-carrier density and magnetic field; see Fig.~\ref{fig:phase_diagram}. Alternatively, superconducting pairing in a two-dimensional electron gas may also be induced by the proximity effect  \cite{Manfra_InAs_MobilityExceeding} with a superconducting film. In that case, there is an additional option of varying $\Delta$ at a fixed value of the carrier density and out-of-plane magnetic field, by changing, e.g., its in-plane component, thus affecting the state of the proximitizing film. 
In either case, as we will show below, an increase of $\Delta$ leads the system through a sequence of topologically-nontrivial superconducting phases ending in a trivial $s$-wave superconductor.

Before proceeding, we clarify the meaning of the topologically different states in the problem at hand. In the limit of vanishing superconductivity, the Bogoliubov--de Gennes (BdG) Hamiltonian describing a superconductor breaks down into the two copies of IQHE Hamiltonians for electrons and holes. 
The IQHE Hamiltonian, in terms of the tenfold symmetry classification  \cite{AltlandZirnbauer, Kitaev2009, RyuSchnyderFurusakiLudwig, EversMirlin, KoziyCrossover}, belongs to class A and is characterized by an integer topological number $\mathds{Z}$. Introduction of the superconducting pairing
brings the
Hamiltonian to class C with the doubled topological index. That enables the initial survival of the topological edge modes, which are composed now of the electron and hole edge states having the same group velocity and propagating in the same direction. 
Therefore, the number and the very presence of the edge modes as a function of the superconducting gap width are not a qualitative question, but rather a quantitative one. Finally, we emphasize that the phases studied in the present work have an even Chern number and therefore do not support Majorana physics. Instead, they possess unfractionalized topological edge modes, which contribute to thermal transport and can be probed via tunneling measurements.

Previous studies of the interplay between superconductivity and Hall physics included analyses of lattice models  \cite{Jain2022helical, Shaffer-HofstadterSC}, effective tight-binding models over Caroli-de Gennes-Matricon bound states in superconducting vortices  \cite{Franz-Tesanovic, Liu-Franz, VafekMelikyanTesanovic}, and continuum models  \cite{DukanAndreevTesanovic1991, TesanovicSacramento, VafekMelikyanTesanovic, McDonald2, McDonald3, McDonald4, McDonald5, McDonald6, McDonald_new, MishmashYazdaniZaletel}. 
A continuum model makes the physical picture more transparent by allowing one to ignore the issue of commensurability between atomic and vortex lattices and the related Harper-Hofstadter phenomenon  \cite{Shaffer-HofstadterSC}. Continuum models also allow one to study the full crossover between perturbed quantum Hall states at $\Delta / (\hbar \omega_c) \ll 1$ and a fully developed superconductivity at $\Delta / (\hbar \omega_c) \gg 1$. Yet, the previous efforts in this direction focused mostly on the single Landau-level limit  \cite{DukanAndreevTesanovic1991, TesanovicSacramento}, reentrant superconductivity at high magnetic fields \cite{McDonald2, McDonald3, McDonald4, McDonald5, McDonald6}, and engineering odd Chern number states by including spin-orbit coupling, applying superlattice potential and/or using an underlying Hamiltonian with the Dirac spectrum.
The investigation of the entire crossover between the quantum Hall and fully developed superconducting regimes, and the possibility of rendering $s$-wave superconductors topological (with even Chern numbers) has received little attention. The purpose of this paper is to fill that gap. In our work, we consider clean systems; the disordered case has recently been studied in Ref.~\onlinecite{Burmi-new}. 

The paper is orgainzed as follows. In Section \ref{sec:model} we introduce our model, and we solve it in Sections~\ref{sec:square-lattice} (square vortex lattice) and 
\ref{sec:traingular-lattice} (triangular vortex  lattice), followed by a discussion in Section~\ref{sec:discussion}. We relegate computational technicalities to Appendixes~\ref{app:translational-properties} and \ref{sec:evaluating_Delta}, analyze the scaling of the matrix elements in Appendix~\ref{app:scaling-Delta}, and explain the subtleties of the Chern number evaluation in a continuum model in Appendix \ref{app:calculating-Chern-continuum-model}. We provide additional results for the bulk properties in Appendix \ref{sec:additional-results}, and we consider the stripe geometry with edge modes in Appendix~\ref{app:stripe-geometry}.

\section{Model and approach}
\label{sec:model}

We start with the model of a two-dimensional electron gas 
with a quadratic dispersion relation 
placed in an out-of-plane uniform magnetic field $B$. The latter can be represented by the vector potential in the Landau gauge $\bA = (-By, 0, 0)$. We assume small effective mass of an electron, which renders the Zeeman term negligible. Its Hamiltonian $H_0 = (-i\bm{\nabla} - e \bA/c)^2 / 2m - E_F$ has energy levels $E_N = (N + 1 / 2) \omega_c - E_F$ with the cyclotron frequency $\omega_c = e B / (m c)$ and well-known Landau-level wavefunctions:
\be \label{Landau_phi}
 \phi_{N,K_x l_B^2} (\br) = e^{i K_x x}\varphi_{N}(y - K_x l_B^{2}) / \sqrt{L_x},
\ee
where $L_x$ is the $x$-dimension of the sample, $\varphi_{N}(y)=\sqrt{1/(2^{N}N!\sqrt{\pi}l_B)} H_{N}(y / l_B)e^{-y^{2}/(2l_B^{2})}$, and Hermite polynomials are denoted as $H_N$. We will use this basis to study both the normal-state and superconducting phases. Note that working in the continuum model without an underlying atomic lattice helps clarify the crossover physics, as it eliminates the need to consider commensurability effects and Hofstadter-related phenomena  \cite{Hofstadter, Harper, Jain2022helical, Shaffer-HofstadterSC}.

Magnetic field is known to be detrimental to superconductivity and suppresses it completely in bulk samples at $B > H_{c2}$. Below the phase transition, $B < H_{c2}$, superconductivity survives in the form of a vortex lattice with a period determined by the magnetic length $l_B = \sqrt{\hbar c/(eB)}$. Specifically, the area of a vortex unit cell equals $\pi l_B^2$.
Note that in the limit of a two-dimensional system, superconducting penetration length diverges  \cite{Pearl-penetration-length} and the magnetic field remains uniform.

On the mean-field level, the superconducting quasiparticle spectrum is described by the Bogoliubov-de Gennes Hamiltonian  \cite{Bogoliubov, deGennes_book}.
It can be written as a matrix
\be \label{BdG}
	\Hbdg = 
	\begin{pmatrix}
		\hat{c}_{\uparrow} \\ \hat{c}_{\downarrow}^\dagger
	\end{pmatrix}^{\dagger}
	 \begin{pmatrix}
		H_0 & \Delta (\br) \\
		\Delta^*(\br) & -H_0^T
	\end{pmatrix}
	\begin{pmatrix}
		\hat{c}_{\uparrow} \\ \hat{c}_{\downarrow}^\dagger
	\end{pmatrix}
\ee
in the Nambu space comprised by spin-up electrons and their time-reversed counterparts, spin-down holes.

Slightly below the critical field $B \lesssim H_{c2}$, one can use an Abrikosov prescription  \cite{Abrikosov_vortex_lattice} for the nascent superconducting order parameter $\Delta(\br)$.  Technically, it is determined by the solution of the Ginzburg-Landau equation with weak nonlinearity and has the form of a vortex lattice. In this regime, the same spatial scale $l_B$ determines both vortex core size and distance between the vortices. The linearized equation for $\Delta(\br)$ is mathematically equivalent to the Schr{\"o}dinger equation of a single Cooper pair in a magnetic field. Therefore, $\Delta(\br)$ can be represented as a linear combination of lowest Landau level wavefunctions of a charge-2e particle. The specific form of the linear combination determines the structure of the vortex lattice. The least-cumbersome case corresponds to a square lattice. We analyze the quasiparticle spectrum for that case before presenting the results for the more realistic triangular lattice.

\section{Square vortex lattice}
\label{sec:square-lattice}
{}
\begin{figure}\includegraphics[width=0.47\textwidth]{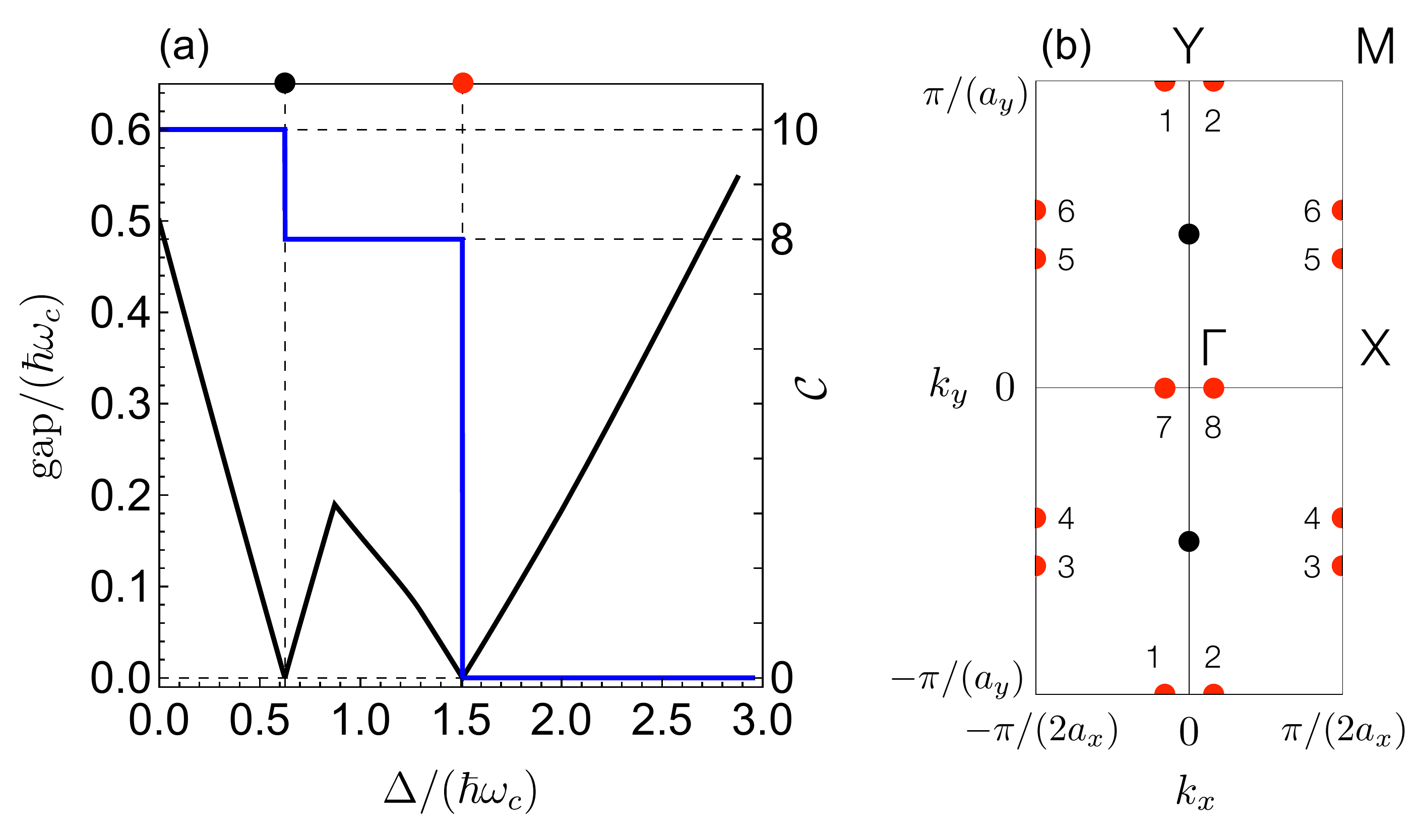}
\caption{Quasiparticle properties in a square vortex lattice. (a) Evolution of the quasiparticle gap (black) and superconducting topological number $\mathcal{C}$ (blue)  as a function of the ratio between the amplitude of the superconducting order parameter $\Delta$ and cyclotron quantum $\hbar\omega_c = e\hbar B / (mc)$. 
We use midgap value $E_F / (\hbar \omega_c) = 5$ corresponding to the filling factor $\nu = 10$ for spin-1/2 electrons.
(b) Brillouin zone of BdG quasiparticles in a square vortex lattice. Positions of the Dirac points in the spectrum at the first (second) gap closure are indicated with black (red) dots. The number of non-equivalent Dirac points equals the jump in the topological index shown in the left panel.
\label{fig:Chern_gap}
}
\end{figure}

In this case, the vortex lattice constants are $a_{x,y} = \sqrt{\pi} l_B$ and $\Delta(\br)$ takes the form  \cite{Abrikosov_vortex_lattice}:
\be \label{Delta_r}
	 \Delta(\br) = 2^{1/4} \Delta \sum_t  \phi_{0, \sqrt{2} t a_y} (\sqrt{2}\br),
\ee
where $\sqrt{2}$ factors account for the doubled charge of a Cooper pair. 
The absolute value $|\Delta(\br)|$ is periodic, while the phase changes non-trivially upon translation by a lattice period, see Appendix~\ref{app:translational-properties}.

Further analysis of BdG Hamiltonian \eqref{BdG} is simplified in the basis of the generalized (magnetic) Bloch wavefunctions of the Landau levels, which  satisfy the boundary conditions that align with the translational properties of $\Delta(\br)$.
For a square lattice this basis is formed by wavefunctions
\be \label{Landau_Fourier_psi}
	\psi_{N, k_x, k_y}^e (\br) = \sqrt{\frac{a_{y}}{L_{y}}} \sum_t e^{i k_y a_y t} \phi_{N, k_x l_B^2 + a_y t} (\br).
\ee
They are parametrized by quasimomentum $\bk = (k_x, k_y)$ belonging to the BdG quasiparticle Brillouin zone (BZ) $k_x \in [0, \pi / a_x]$ and $k_y \in [0, 2\pi / a_y]$ {--- in real space, the quasiparticle unit cell is twice as large as the vortex unit cell  \cite{DukanTesanovic, Franz-Tesanovic, Liu-Franz}.} A basis wavefunction \eqref{Landau_Fourier_psi} can be viewed as a superposition of a set of functions formed by center-of-orbit translations, $K_xl_B^2\to K_xl_B^2+a_yt$, of the Landau wavefunctions (1); the respective amplitudes in the superposition are $e^{ik_ya_yt}$.

Applying time-reversal operation, one gets eigenfunctions of the hole Hamiltonian $-H_0^T$ in the form $\psi_{N, \bk}^h = (\psi_{N, -\bk}^e)^*$. 
Wavefunctions \eqref{Landau_Fourier_psi} 
have a generalized periodicity property, 
acquiring an additional field-dependent phase upon a translation by a vortex lattice period, see Appendix~\ref{app:translational-properties}. As a result, the product $\psi^{e} (\psi^{h})^* $ transforms under translations the same way as  $\Delta(\br)$. 
Therefore, the matrix elements of $\Delta(\br)$ are diagonal in this basis  \cite{McDonald2, DukanAndreevTesanovic1991, McDonald4, McDonald6}:
\be \label{Delta_Landau_basis}
	\left\langle \psi_{N,\bk}^e |\Delta(\br) | \psi_{M,\bk'}^h \right\rangle = \delta_{\bk, \bk'} \Delta_{NM} (\bk),
\ee
with the coefficients $\Delta_{NM}(\bk)$
\be \label{Delta_NM_k}
    \Delta_{NM}(\bk) = \Delta  \sum_{t = -\infty}^{\infty} c_0 e^{2 i k_y a_y t} \varphi_{M + N}(2t a_y + 2k_x l^2),
\ee
where $c_0 = (-1)^N 2^{-(M+N)/2} \sqrt{C_{M+N}^M}$ and $C_{M+N}^M = (M+N)!/(M!N!)$ is the binomial coefficient. 
Details of the derivation of Eq.~(\ref{Delta_NM_k}) can be found in Refs.~\onlinecite{McDonald4} and in Appendix B.

Substitution of Eq.~\eqref{Delta_Landau_basis} for $\Delta$ and of Landau level energies $E_N = \omega_c (1 / 2 + N) - E_F$ for $H_0$ into \eqref{BdG} gives BdG Bloch Hamiltonian $H_{\text{BdG}}$ in the basis \eqref{Landau_Fourier_psi}. It is  parametrized by a two-dimensional quasimomentum $\bk$. With a cutoff for LL index $N_{\max}$, both for electrons and holes, it becomes a $\bk$-dependent $2N_{\max} \times 2N_{\max}$ matrix. 

Diagonalizing it numerically with a cutoff $N_{\max} = 30$, we find the quasiparticle energy spectrum. We consider even filling factors $\nu$ using the Fermi energy $E_F = \hbar \omega_c (1/2 + \nu/2)$, which corresponds to a midgap point between Landau levels at $\Delta = 0$. Apart from the spectrum, we study the topological index $\mathcal{C}$ of the system, 
which gives the number of protected edge modes. 
According to the generalized TKNN (Thouless-Kohmoto-Nightingale-den Nijs) formula  \cite{TKNN, Jain2022helical}, it equals the sum of superconducting Chern numbers of occupied bands:
\be \label{TKNN}
    \mathcal{C}_i = -\frac{i}{2\pi} \int_{\text{BZ}} d^2\bk\, \mathcal{F}_{ii}(\bk),
\ee
where $\mathcal{F}_{ii}(\bk)$ is the Berry curvature of the $i$th band:
\be \label{F-BerryCurvature}
    \mathcal{F}_{ii}(\bk) = \epsilon_{lm} \partial_{k_l} \left\langle u_{i,\bk} | \partial_{k_m} | u_{j,\bk}\right\rangle.
\ee
Equation \eqref{F-BerryCurvature} is written in terms of the Bloch wavevector $u_{i, \bk}$, related to the full wavefunction by the plane-wave factor $\psi_i(\bk) = e^{i\bk \br} u_{i, \bk}$. 
Note that electrons and holes belong to different subspaces of the BdG Hilbert space, thus there is no cross electron-hole contribution to the scalar product in Eq.~\eqref{TKNN}. Summation over all occupied bands of a BdG Hamiltonian for a continuum model encounters a difficulty because the spectrum is unbounded from below. We describe how this difficulty can be resolved in Appendix~\ref{app:regularization}. 
For a numerical calculation, we use a $32\times32$ mesh in the Brillouin zone.

\textit{Topological transitions.}
In the absence of superconductivity, $\Delta=0$, BdG spectrum consists of electron and hole Landau levels.  The quasiparticle gap is of the order of  $\omega_c$ as it is determined by the energy separation between the Fermi level and the closest Landau level. Upon increasing $\Delta$ from zero, electrons and holes form hybridized energy bands. Due to the spatial dependence of $\Delta(\br)$, the formed bands acquire dispersion, thus reducing the energy gap. At some value of $\Delta$, the gap closes  \cite{VafekMelikyanTesanovic}. 
Remarkably, it then undergoes {a series of revivals
to the values $\sim 0.3 \hbar\omega_c$.} This is illustrated for the filling factor $\nu=10$ by a black line in Fig.~\ref{fig:Chern_gap}(a). 

Each gap closure is accompanied by a change in the topological number $\mathcal{C}$ as shown in blue in Fig.~\ref{fig:Chern_gap}(a). Notably, $\mathcal{C}$ can change by  different and relatively large integers at each gap closure [e.g. by $8$ in the second gap closure of Fig.~\ref{fig:Chern_gap}(a)]. Further inspection reveals that the gap closes through the occurrence of multiple Dirac points  \cite{TesanovicSacramento} in the Brillouin zone at the same value of $\Delta/\omega_c$; see Fig.~\ref{fig:Chern_gap}(b). The number of Dirac points corresponds to the jump of the topological index at the transition, while their position varies. We also confirm bulk-boundary correspondence in our system by showing that the number of edge modes in the stripe geometry is indeed given by $\mathcal{C}$, see Appendix~\ref{app:stripe-geometry}.

The topological number eventually reaches zero at large $\Delta$, and no edge states are left after the last gap closing. {As will be shown below, this behavior holds for larger filling factors as well; see Fig.~\ref{fig:Cherns_triangular} for the case of the triangular vortex lattice.} The result suggests that indeed the topological edge modes survive the onset of superconducting order, but eventually disappear with the increase of $\Delta$ after several consecutive gap closures in the bulk.

After the last gap closing, the gap grows approximately linearly; see Fig. \ref{fig:Chern_gap}.
This behavior is limited to the regime of nascent superconductivity where the coherence length $\xi \equiv \hbar v_F / \Delta \gtrsim l_B$; this implies $\Delta/\omega_c \lesssim \sqrt{\nu}$. 
Under these conditions, the spatial scale of CdGM states is still of the order of intervortex distance, and CdGM states on the vortex lattice are still strongly hybridized.
At larger values of $\Delta$, beyond the studied regime, the bound states localize in individual vortices, and the spectral gap becomes associated with the energy of the lowest CdGM level, $\sim \Delta^2 / E_F$. Therefore, the spectral gap continues to grow  with the increase of $\Delta$.

\section{Triangular lattice}
\label{sec:traingular-lattice}
\begin{figure}
	\includegraphics[width=0.425\textwidth]{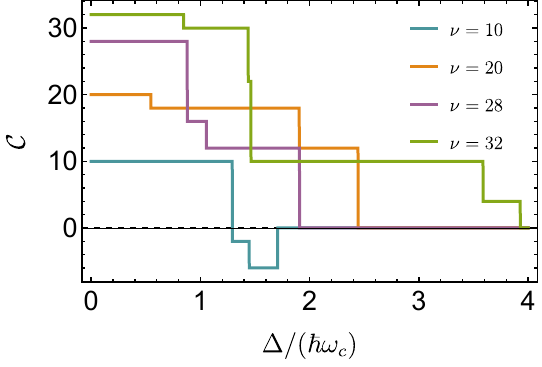}
	\caption{
 Evolution of the superconducting topological number in the triangular vortex lattice as a function of the ratio between the amplitude of the superconducting order parameter $\Delta$ and cyclotron energy $\hbar \omega_c$ for different even-integer filling factors $\nu$. 
 }
	\label{fig:Cherns_triangular}
\end{figure}

In the previous paragraph, we studied the simplest example of a square vortex lattice, allowing for the most compact analytical treatment. On the other hand, in most cases the energetically favorable configuration of the superconducting vortices is a triangular lattice. The calculations generalize straightforwardly to this case producing  similar results.
The spatial profile of $\Delta$ is given in this case by  \cite{Abrikosov_vortex_lattice}:
\be \label{Delta_r_triang}
	 \Delta(\br) = 2^{1/4} \Delta_0 \sum_t \exp(-i \pi t^2 /2) \phi_{0, \sqrt{2} t a_y} (\sqrt{2}\br).
\ee
{The generalization of Eqs.~\eqref{Landau_Fourier_psi}--\eqref{Delta_NM_k} for the gap structure given by Eq.~\eqref{Delta_r_triang} is presented in the Appendix A.} The results of the numerical calculation are shown in Fig.~\ref{fig:Cherns_triangular} for several filling factors. At any $\nu$, the Chern number drops to zero in several consecutive jumps, similar to the case of the square lattice.

For the considered lattices, the superconducting Chern number changes by an even integer at each gap closing. We believe this is a salient feature of the Bravais vortex lattice structure with single magnetic flux quantum  $h/(2e)$ per unit cell.
A more complicated vortex arrangement (e.g., two vortices per unit cell) may support additional phases, bearing an odd Chern number, leading to the possibility of Majorana physics at the edges  \cite{ZocherRosenow, MishmashYazdaniZaletel, McDonald_new}.

\section{Discussion} 
\label{sec:discussion}

The main conclusion of our work is that a two-dimensional $s$-wave superconductor in an out-of-plane magnetic field undergoes a number of phase transitions between topologically-nontrivial states on its way to a conventional, trivial superconducting state carrying well-separated vortices in the structure of its order parameter. This evolution occurs in the process of increasing $\Delta/\hbar\omega_c$, and each topological transition is accompanied by a change in the number of gapless edge modes, reaching zero in the trivial state. On the basis of perturbation theory in $\Delta$, we expect the first transition to occur once the width ${\rm max}_{\bk}|\Delta_{NN}(\bk)|$ of the band developing from $N$-th Landau level becomes of the order of the distance between the levels, $\hbar\omega_c$. The large-$\nu$ asymptote of the band width is ${\rm max}_{\bk}|\Delta_{NN}(\bk)|\sim\Delta/\nu^{1/4}$, see Appendix~\ref{app:scaling-Delta}. Condition ${\rm max}_{\bk}|\Delta_{NN}(\bk)|\sim\hbar\omega_c$ then yields $\Delta/\hbar\omega_c\sim\nu^{1/4}$ for the first transition. 
One may expect that, upon a cascade of transitions, the trivial state is finally reached once the vortices are well separated, {\it i.e.} under condition $l_B\gtrsim\xi$. It can be re-written as $\Delta/\hbar\omega_c\gtrsim \sqrt{\nu}$. 
Results for the sequence of transitions in the triangular vortex lattice calculated at different $\nu$ are roughly consistent with these expectations, see Fig.~\ref{fig:Cherns_triangular}.

The electron pairing field may result from interaction within the two-dimensional system, or induced by proximity to a superconductor. In the former case, 
$\Delta$ is determined by the BCS self-consistency equation, and therefore depends on the magnetic field $B$  \cite{ GorkovHc2, HelfandWerthamer}. 
Upon the decrease of $B$ along one of the fixed-$\nu$ lines, cf. Fig.~\ref{fig:phase_diagram}, 
the gap parameter raises from zero, once $B$ is lowered below its critical value,  $H_{c2}$~\footnote{For a two-dimensional system, it is safe to assume the type-II superconductivity.}. This poses the question of how far below the $H_{c2}$ field the edge modes survive. The $\Delta(B)$ dependence at $B$ close to $H_{c2}$ can be approximated~ \cite{Abrikosov_vortex_lattice} as $\Delta (B)=\Delta_0 (1 - B / H_{c2})^{1/2}$. Here $\Delta_0$, up to a numerical factor $\sim 1$, is the gap value at $B=0$ and fixed density. We use this approximation for $\Delta(B)$ and switch from fixed density to fixed filling factor  $\nu\approx 2E_F/\hbar\omega_c$ to find that the condition for the first quasiparticle gap closing, $\Delta(B)\sim\hbar\omega_c(B)[\nu(B)]^{1/4}$, is reached at field $B_{\text{top}}^\prime=H_{c2} - \delta H_{\text{top}}^\prime$, with $\delta H_{\text{top}}^\prime / H_{c2} \sim 1/\sqrt\nu$. 
A similar analysis of the transition to the topologically-trivial state leads us to the estimate $B_{\text{top}}^{\prime\prime}=H_{c2} - \delta H_{\text{top}}^{\prime\prime}$, with $\delta H_\text{top}^{\prime\prime}\sim a H_{c2}$ and $a\lesssim 1$. 

The variation of $H_{c2}$ with the carrier density is model-dependent. In the case of a short-range pairing interaction the variation is weak and, using Ref.~\onlinecite{Miyake}, we estimate $H_{c2}\sim(\Phi_0/\pi\hbar^2)mE_b$. Here $E_b$ is the binding energy of a two-particle state, and $m$ is the particle mass.

Regardless of microscopic details, at very weak field $H\ll H_{c2}$ (where $\Delta/\hbar \omega_c\gg 1$), the $s$-wave superconducting state with well-separated vortices is topologically trivial, because it is adiabatically connected  \cite{Leggett_ModernTrends, Leggett_ParisJournal, Randeria_review} to the BEC limit, 
a superfluid of tightly bound Cooper pairs in the presence of vortices, which does not support any low-lying fermionic excitations either in the bulk or at the edge. Indeed, in the self-consistent mean field theory  \cite{Randeria_vortices_BECBCS}, the number of CdGM levels in a vortex decreases when electron concentration is lowered at fixed interaction strength, and the fermionic gap never closes. The energy of the lowest CdGM state is of the order of $ \Delta^2 / E_F$ and approaches Cooper pair binding energy $E_b$ when $\Delta$ reaches its zero-field value given by $\sqrt{2 E_F E_b}$  \cite{Miyake}.

At high filling factors, the 
sequence of the topological transitions starts in a narrow region $\delta H_{\text{top}}^\prime$ near $H_{c2}$,
see Fig.~\ref{fig:phase_diagram}. It is instructive to compare  $\delta H_{\text{top}}^\prime$ with the width $\delta H_{\text{fluct}}$ of the critical fluctuations domain around the quantum phase transition at $B=H_{c2}$  \cite{LarkinVarlamov}. Width $\delta H_{\text{fluct}}$ can be estimated from a comparison of the quantum fluctuations contribution to the magnetic susceptibility with the susceptibility jump in the
mean-field theory of the phase transition~ \cite{GalitskiLarkin}. This comparison for a two-dimensional superconductor, along with the relation $\nu = 2 E_F/\hbar\omega_c$, yields $\delta H_{\text{fluct}} / H_{c2} \sim 1/\sqrt{\nu}$.
As follows from the previous paragraph, the start of the  sequence of topological transitions falls into the region of critical fluctuations at any $\nu$. 

In a hypothetical case of superconductivity induced in a two-dimensional electron gas by a proximitizing material with a very high $H_{c2}$, one may use variation of the magnetic field to tune $\hbar\omega_c$ of the electron gas without affecting the induced value of $\Delta$. The main advantage of a hybrid structure is the possibility of tuning through the cascade of topological transitions from one with the highest Chern number down to the trivial state in the absence of the quantum fluctuations of the vortex lattice. Furthermore, the vortex lattice structure induced from outside does not melt, even at small $\nu$ in the two-dimensional electron gas. Lastly, the cascade of transitions even at relatively small values of $\nu\sim 8\div12$ occurs in a fairly wide range of fields in which $\hbar\omega_c$ is reduced from $\sim 2\Delta$ to $\sim 0.5\Delta$. 

A prerequisite for observing the cascade of topological transitions common for the intrinsic and proximitized superconductors is a 
large elastic electron relaxation time, $\tau\gg\omega_c^{-1}$. It is needed in order to have well-resolved Landau levels and robust edge states at $\hbar\omega_c\sim\Delta$, leading to the requirement of a clean-superconductor limit, $\Delta\tau\gg 1$. Some graphene-based structures exhibiting superconductivity are believed to be in this regime~ \cite{ParkNature}. In an InAs/Al structure, the induced superconducting gap is constrained by the gap value $\Delta \sim 3 \cdot 10^{-4} \,  \text{eV}$ in the Al layer, and the condition $\hbar\omega_c\sim\Delta$ is reached at $B\sim 0.1 \, \text{T}$ (we used here the InAs effective electron mass $m^* = 0.026 m_e$). At a record electron mobility~ \cite{Manfra_InAs_MobilityExceeding} of $\mu \sim 10^5 \text{cm}^2 \, / (\text{V} \cdot \text{s})$ we estimate $\Delta\tau\sim 10^4$. Achieving $H_{c2}\gtrsim 0.1 \, \text{T}$ in the Al layer, however, may pose a challenge: it requires a very short electron mean free path of $\sim 10\,\text{nm}$ in that layer. Topological transitions can be probed either by measuring spin and thermal Hall conductivity at ultra-low temperatures or by probing edge modes by tunneling spectroscopy.

Finally, in this work we focused on the single-particle properties of intrinsic or proximitized two-dimensional superconductors subject to an out-of-plane magnetic field. A separate and interesting question is about the evolution with $\nu$ of the Hall resistivity of an intrinsic superconductor. This question will be addressed in our future work.

\section{Acknowledgments}
We acknowledge many instructive discussions with colleagues including Cenke Xu and Max Metliski. This work was supported at Yale University  by NSF Grant No. DMR-2410182 and by the Office of Naval Research (ONR) under Award No. N00014-22-1-2764.

\nocite{McDonald6, Hofstadter, Harper, Wen-Zee, Shaffer-HofstadterSC, Fukui-Berry}

\appendix 

\section{Translational properties of the superconducting pairing $\Delta(\br)$ and the generalized Bloch wavefunctions}
\label{app:translational-properties}

In this Section we present translational properties of $\Delta(\br)$ and BdG quasiparticle wavefunctions given by Eq.~\eqref{Landau_Fourier_psi} of the main text and show their consistency with each other. Then, we demonstrate the generalization to the case of the triangular vortex lattice. 

\subsection{Square lattice}
In this appendix, we consider a square vortex lattice with the unit vectors $a_x = a_y = \sqrt{\pi} l_B$ corresponding to one superconducting flux quantum per vortex. In this case, we used Abrikosov's ansatz for the superconducting pairing $\Delta(\br)$ given by Eq.~\eqref{Delta_r} of the main part. In the chosen gauge, it has the following properties under translations by the unit vectors:
\begin{subequations}
\label{Delta_translations}
\begin{align}
    \Delta(x + a_x, y) &= \Delta(x, y),\\ 
    \Delta(x, y + a_y) &= e^{2 \pi i x / a_x} \Delta(x,y) .
\end{align}
\end{subequations}
At the same time, the wavefunctions \eqref{Landau_Fourier_psi} have the following properties:
\begin{subequations}
\label{psi_translations}
\begin{align}
    & \psi_{N, \bk}^{e(h)} (x + 2 a_x, y) = e^{2 i k_x a_x} \psi_{N, \bk}^{e(h)} (x, a_y), \\ 
    & \psi_{N, \bk}^{e(h)} (x, y + a_y) = e^{\pm i \pi x / a_x} e^{i k_y a_y} \psi_{N, \bk}^{e(h)} (x, a_y).
\end{align}
\end{subequations}
Note that \eqref{Landau_Fourier_psi} do not have well-defined translational properties under $x \rightarrow x + a_x$ but do transform as Bloch wavefunctions under $x$-translation by $2 a_x$. Thus, the BdG quasiparticle real-space unit cell is twice as large as the vortex unit cell  \cite{Liu-Franz}, which is related to the fact that electron and hole charge is two times smaller than the charge of a Cooper pair. Consequently, in momentum space, BdG quasiparticles have the Brillouin zone defined by intervals $k_x \in [0, \pi / a_x]$ and $k_y \in [0, 2\pi / a_y]$, which is two times smaller than the naive Brillouin zone of a vortex lattice.

Quasiperiodicity properties \eqref{Delta_translations} and \eqref{psi_translations} are consistent --- as written in the main part, the product $\psi^{e} (\psi^{h})^* $ transforms in the same way as  $\Delta(\br)$. That's why the Hamiltonian in the basis \eqref{Landau_Fourier_psi} is rendered diagonal in momentum space; see Eqs.~\eqref{Delta_Landau_basis} and \eqref{Delta_NM_k} of the main text.

\subsection{Triangular lattice}

In the triangular vortex lattice case, the Abrikosov expression for $\Delta(\br)$ is modified by an additional factor $e^{-i\pi t^2/2}$ under the sum over $t$, see Eq.~\eqref{Delta_r_triang} of the main part. Correspondingly, the translational properties now read:
\begin{subequations}
\label{Delta_translations_triangular}
\begin{align}
    & \Delta(x + a_x, y) =  \Delta(x, y),\\ 
    & \Delta(x + a_x / 2, y + a_y) = -i e^{2 \pi i x / a_x} \Delta(x,y) .
\end{align}
\end{subequations}
Note that now the principal translational vectors $(a_x, 0)$ and $(a_x/2, a_y)$ corresponds to the triangular lattice. 

To account for the modified pairing $\Delta(\br)$, the expression for the basis of wavefunctions should also be modified. Specifically, instead of Eq.~\eqref{Landau_Fourier_psi} one should use for electrons:
\be \label{Landau_Fourier_psi_traingular}
	\psi_{N, k_x, k_y}^e (\br) = \sqrt{\frac{a_{y}}{L_{y}}} \sum_t e^{i k_y a_y t} e^{-i \pi t^2 / 4} \phi_{N, k_x l_B^2 + a_y t} (\br).
\ee
The expression for holes is still obtained by the operation $\psi_{N, \bk}^h = (\psi_{N, -\bk}^e)^*$. Translational properties of \eqref{Landau_Fourier_psi_traingular} now read:
\begin{subequations}
\label{psi_translations_triangular}
\begin{align} 
    & \psi_{N, \bk}^{e(h)} (x + 2 a_x, y) = e^{2 i k_x a_x} \psi_{N, \bk}^{e(h)} (x, a_y), \\ 
    & \psi_{N, \bk}^{e(h)} (x + a_x / 2, y + a_y) = e^{i k_x a_x/2 + i k_y a_y \pm i \pi /4 \pm i \pi x / a_x} \nonumber \\
    & \qquad \qquad \qquad \qquad \qquad \qquad \qquad\times \psi_{N, \bk}^{e(h)} (x, a_y).
\end{align}
\end{subequations}
As in the square lattice case, the unit cell for BdG quasiparticles is twice as large as the vortex unit cell. 
It is straightforward to check that $\psi^{e} (\psi^{h})^* $ again transforms the same way as $\Delta(\br)$ so that \eqref{Landau_Fourier_psi_traingular} is the correct basis for the triangular vortex lattice.

\section{Calculation of the matrix elements of $\Delta(\br)$}
\label{sec:evaluating_Delta}

The matrix elements of the vortex-state $\Delta(\br)$ in the basis of Landau levels were obtained in Refs.~\onlinecite{McDonald2, McDonald3, McDonald4, McDonald5, McDonald6}. In this Appendix we outline the derivation procedure for completeness. The key step is obtaining a useful identity  \cite{McDonald2} for the Landau level wavefunctions, see Eq.~\eqref{Landau_phi} in the main text:
\begin{align}
\label{LL_identity}
    & \phi_{N,Y} (\br_1) \phi_{M,Y'}(\br_2)   
    \\ \nonumber
    & \quad
    = \sum_j \mathcal{B}_j^{N,M} \phi_{j, Y_c}^R ([\br_1 + \br_2]/2) \phi_{N+M-j, Y_r}^r (\br_1 - \br_2),
\end{align}
where $\phi^{R(r)}$ differ from $\phi$ by a substitution $l_B \rightarrow (\sqrt{2})^{\mp 1} l_B$ correspondingly and 
\be \label{useful_identity}
	\mathcal{B}_j^{N,M}  = \sum_{m=0}^j \frac{(-1)^{M-m}}{2^{N+M}} \sqrt{C_j^m C_M^m C_N^{j-m} C_{N+M-j}^{M-m}}
\ee
with the the binomial coefficients $C_n^m = n! / (m! (n-m)!)$. To obtain \eqref{LL_identity}, one starts with the Hamiltonian of two non-interacting particles in magnetic field in the Landau gauge, $H_{1,2} = H_0(\br_1) + H_0(\br_2)$, where $H_0$ is given in the main text. Next, consider the basis change to the center-of-mass $\bR = (\br_1 + \br_2)/2$ and relative $\br = \br_1 - \br_2$ coordinates. Due to the linearity of the Landau gauged vector potential in the spatial coordinates, one can rewrite $H_{1,2} = H_R(\bR) + H_r(\br)$, where $H_R (H_r)$ have the same form as $H_0$ but with double (half) mass and charge. The ground state of the system corresponds to both particles residing in the lowest Landau level ($N=M=0$) or, equivalently, occupation of the lowest Landau level both for center-of-mass and relative coordinates. Then, Eq.~\eqref{useful_identity} for $N=M=0$ can be obtained via straightforward check. General form of \eqref{LL_identity} is obtained by using the algebra of creation/annihilation operators $a_{1(2)} = (l_B / \sqrt{2}) (\pi_{1(2),x}) - \pi_{1(2),y}) $ with the generalized momentum $\bm{\pi}_{1(2)} = i \bm{\nabla}_{1(2)} + (e/c) \bA_{1(2)}$ while $a_R = (a_1 + a_2)/\sqrt{2}$ and $a_r = (a_1 - a_2)/\sqrt{2}$. 

The superconducting vortex lattice of a general profile in the vicinity of the transition is given by a generalization of Eqs.~\eqref{Delta_r} and \eqref{Delta_r_triang}:
\be \label{Delta_general}
	\Delta(\br) =  \sum_t \Delta_t \phi_{0,\sqrt{2} t a_y} (\sqrt{2} \br)
\ee
with $\Delta_t$ being some periodic function of $t$. 

The first step in the calculation is to consider matrix elements $\langle \phi_{N,K_x l_B^2}|\Delta(r)|\phi_{M, K_x'}^*\rangle$. With the substitution $K_x = k_x + a_y t / l_B^2$, where $k_x \in [0,\pi / a_x]$ and integer $t$, they can be viewed as a `mixed' real-momentum space representation, where $k_x$ parametrizes $x$-component of momentum, while $t$ parametrizes $y$-coordinate of the unit cell in the real space. Performing a (generalized) Fourier transform of $\phi_{N,k_x l_B^2 + a_y t}$ in the parameter $t$ leads exactly to the correct bases \eqref{Landau_Fourier_psi} and \eqref{Landau_Fourier_psi_traingular} that make the Hamiltonian diagonal in $\bk$. For the matrix element calculation, one substitutes \eqref{Delta_general} into $\langle \phi_{N,k_x l_B^2 + a_y t}|\Delta(r)|\phi_{M, k_x' l_B^2 + a_y t'}^*\rangle$ and uses the identity \eqref{LL_identity} and orthogonality properties of the oscillator wavefunctions.

Finally, one substitutes \eqref{Landau_Fourier_psi} or \eqref{Landau_Fourier_psi_traingular} in the definition $\left\langle \psi_{N,\bk}^e |\Delta(\br) | \psi_{M,\bk'}^h \right\rangle$ and performs sums over $t$ and $t'$ using the knowledge of $\Delta_t$ in a specific vortex configuration. In this way, one obtains Eqs.~\eqref{Delta_Landau_basis}--\eqref{Delta_NM_k} for the square lattice and the following expression for the triangular lattice:
\be \label{Delta_NM_k_tr}
    \Delta_{NM}(\bk) = c_0 \sum_{t = -\infty}^{\infty} e^{2 i k_y a_y t - i \pi t^2 / 2} \varphi_{M + N}(2t a_y + 2k_x l^2).
\ee
For convenience, we repeat the square-lattice expression, Eq.~\eqref{Delta_Landau_basis} and the value of $c_0$ below, see Eq.~\eqref{DeltaMN-SM}. Expressions \eqref{Delta_NM_k_tr} and \eqref{DeltaMN-SM} define the off-diagonal part of the BdG Hamiltonian in momentum space while the main-diagonal part is given by Landau level energies. 

\section{Scaling of the matrix elements $\Delta_{N, M} (\bk)$ at large $N, M$}
\label{app:scaling-Delta}

\begin{figure}[h!]
	\includegraphics[width=0.45\textwidth]{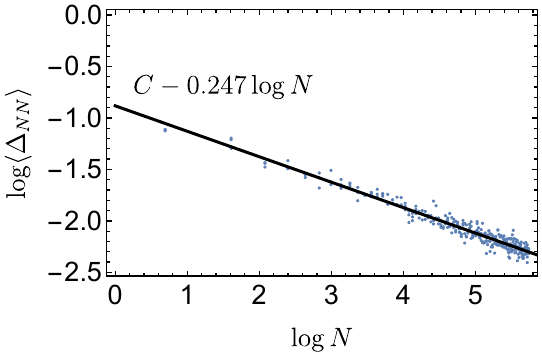}
	\caption{Scaling of the typical matrix elements $\Delta_{N, M} (\bk)$ at large $N$. Plotted is the logarithm of the mean of absolute values of $\Delta_{N,N} (\bk)$ evaluated at 100 random points in the BdG Brillouin zone versus the logarithm of the Landau level index $N$. The dependence is fitted well with a linear function of the slope $-1/4$ consistent with the analytical reasoning. }
	\label{fig:Delta-scaling}
\end{figure}
In this appendix, we analyze the behavior of the matrix elements of the superconducting pairing $\Delta_{N, M} (\bk)$ in the limit of large Landau level indices $M$ and $N$. Here $\Delta_{M, N} (\bk)$ is evaluated in the basis of Landau level wavefunctions; see Eq.~\eqref{Delta_r} of the main text for the definition in the square vortex lattice case. The explicit calculation leads to  Eq.~\eqref{Delta_Landau_basis} of the main part, which we repeat here for convenience:
\be \label{DeltaMN-SM}
    \Delta_{NM}(\bk) = \Delta \cdot c_0 \sum_{t = -\infty}^{\infty}  e^{2 i k_y a_y t} \varphi_{M + N}(t a_y + k_x l^2),
\ee
where
\be
    c_0 = (-1)^N 2^{-(M+N)/2} \sqrt{C_{M+N}^M} .
\ee
For simplicity, we will put $M = N$ in the following. The asymptotics of the coefficient $c_0$ is straightforwardly obtained from Stirling's asymptotic approximation for factorials, $c_0 \sim N^{-1/4}$. To estimate the scaling of the rest of the expression given by sum over $t$, we recall that the $\varphi_{N}(x)$ are oscillator wavefunctions, which can be obtained quasiclassically at large $N$. Their spacial scale is given by the cyclotron radius $r_c = \sqrt{N} l_B$, so that the typical value of a properly normalized $\varphi_{N}(x)$ is proportional to $N^{-1/4}$. On the other hand, \eqref{DeltaMN-SM} implies that the sum is cut at $|t| \sim r_c / l_B = \sqrt{N}$ elements. Assuming their complex phase as random for a general $\bk$, we use the central limit theorem arriving at the scaling $\sim N^{-1/4} N^{1/4} = 1$. Then, the typical matrix elements scale as $c_0$:
\be \label{Delta-scaling}
	|\Delta_{NM}(\bk)| \sim \Delta N^{-1/4}.
\ee

To check this result numerically, we sample 100 values of $\Delta_{N, N } (\bk)$ at random points in the BdG Brillouin zone and take the mean of their absolute values. Doing the procedure for $N = 1 \dots 300$, we arrive at a trend which confirms Eq.~\eqref{Delta-scaling}. The results are presented in Fig.~\ref{fig:Delta-scaling} in the logarithmic scale. The same scaling is expected to hold for the triangular vortex lattice, and we confirmed this result numerically.

\section{Calculation of topological number in the continuum model}
\label{app:calculating-Chern-continuum-model}

In this appendix, we present details of the calculation of superconducting topological number calculation. It is generally given by the sum of Chern numbers of occupied bands with the straightforward generalization of the TKNN formula  \cite{TKNN} to the BdG wavefunctions. In the first part, Appendix~\ref{app:regularization}, we use a trick that allows summation over an infinite number of occupied bands in the continuum model. In the second part, Appendix~\ref{sec:Chern-analytical}, we show how the numerical calculation can be simplified if one does a part of the calculation analytically. 

\subsection{Regularizing an infinite sequence of bands}
\label{app:regularization}

\begin{figure}[h]
	\includegraphics[width=0.4\textwidth]{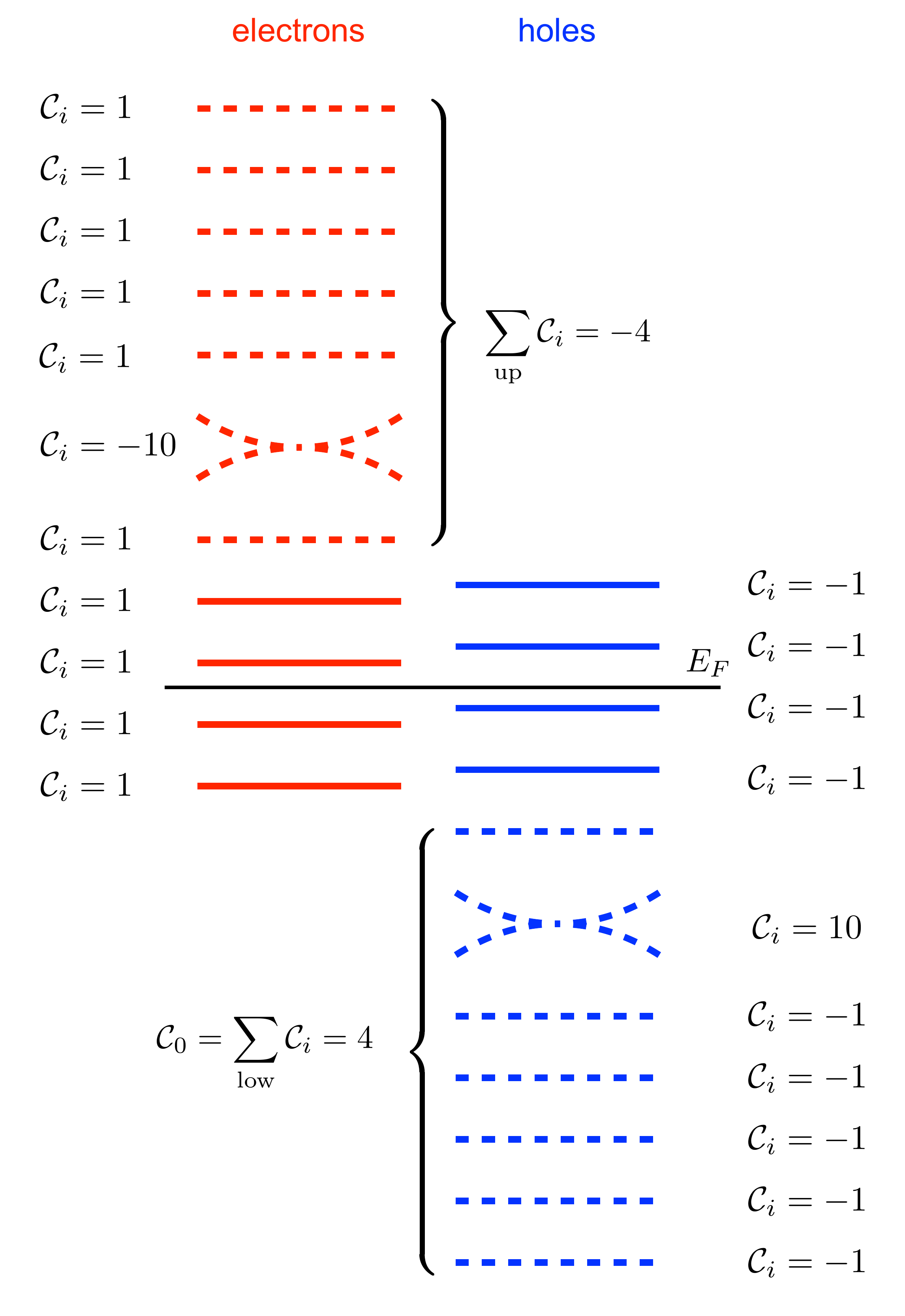}
	\caption{Regularization of the continuum model with an atomic lattice. The scheme represents electron (red) and hole (blue) Landau levels at $\Delta = 0$. In the continuum model, the hole spectrum is unbounded from below, while in the regularized model the number of bands is finite. Bands that are significantly coupled by $\Delta$ (solid lines) are approximated well by a continuum model, but the (constant) contribution of low-lying holes (blue dashed) should be included in the calculation of the total Chern number of occupied bands (topological number). 
}
	\label{fig:LL_Chern_regularization}
\end{figure}

As mentioned in the main part, computation of the BdG topological index given by the total Chern number of occupied bands encounters a hurdle for a continuum model because the spectrum is unbounded from below. Indeed, even in the absence of $\Delta$, the hole block of the BdG Hamiltonian $-\hat{H}_0^T$ leads to the sequence of Landau levels of holes that extends to negative energies without a bound. To deal with this issue, one can consider a  regularization in the form of a square atomic lattice model with a rational flux through a unit cell. In this case, the tight-binding model for electrons  \cite{Hofstadter, Harper} (Harper-Hofstadter model) possesses a finite number of bands, see Fig.~\ref{fig:LL_Chern_regularization}. Note that the total Chern number computed over the whole spectrum (both occupied and unoccupied bands) is zero. For the lowest bands, the continuum approximation is valid and wavefunctions \eqref{Landau_phi} can be used. Higher bands have the same Chern number except for the special band or pair of bands in the middle of the spectrum that has a large Chern number of the opposite sign \cite{Wen-Zee}. 
The spectrum of the hole Hamiltonian $-\hat{H}_0^T$ is obtained by applying time-reversal operation and flipping the energy sign. Thus, there is also a finite number of hole bands with opposite Chern numbers as illustrated in Fig.~\ref{fig:LL_Chern_regularization}.
The BdG Hamiltonian now has a finite number of occupied bands and their total Chern number is well-defined. At $\Delta = 0$, its value is doubled with respect to the TKNN topological index of the electronic Hamiltonian $\hat{H}_0$, as expected. 

The sum over occupied bands remains finite and thus well-defined in the presence of a non-zero $\Delta$ enabling the calculation of the superconducting topological index. Fortunately, one does not have to resort to the full-fledged theory of Hofstadter superconductivity  \cite{Shaffer-HofstadterSC} because strong intermixing and gap closures happen only for the bands that have time-reversed partners with close enough energy, $\delta E \lesssim \Delta$, see the solid lines in Fig.~\ref{fig:LL_Chern_regularization}. Those are low-lying energy bands that can be studied within the continuum model with a certain cutoff $N_{\max} \gg E_F / \omega_c$ for the Landau level index of both electrons and holes. Other bands are considered to be decoupled and approximated by the unperturbed LLs at zero pairing field $\Delta$ (dashed lines in Fig.~\ref{fig:LL_Chern_regularization}). 
Disregarding $\Delta$ for higher LL bands is particularly justified for the computation of the topological number as the Chern number of a band cannot change without a gap closure to another band.
However, when evaluating the total Chern number of the occupied bands, it is important to add the constant contribution $\mathcal{C}_0$ of the low-lying holes (dashed blue). One can infer that $\mathcal{C}_0 = N_{\max}$ by noting that at $\Delta = 0$ the total Chern number of the complementary set of the hole states (solid blue) is $-\mathcal{C}_0$ in the lattice model; the latter are related to the low-lying electrons (solid red) by a time-reversal operation, which flips the Chern number.
To summarize, the accumulated Chern number of the occupied bands can be evaluated in the continuum model by imposing a cutoff $N_{\max}$ for maximal electron and hole LL index and adding an additional contribution $\mathcal{C}_0 = N_{\max}$ to the result:
\be
	\mathcal{C} = \mathcal{C}_0 + \sum_{\text{occup}} \mathcal{C}_i .
\ee

\subsection{Superconducting Chern number calculation}
\label{sec:Chern-analytical}

Chern numbers of the Bogoliubov--de Gennes Hamiltonian are given by the superconducting TKNN formula; see Eq.~\eqref{TKNN}.
It can be effectively computed numerically by dividing the Brillouin zone into small plaquettes and summing the following expression over them \cite{Fukui-Berry}:
\begin{align} \label{Fukui_app}
\mathcal{C} &= \sum_{\square}\arg\Big(\left\langle u_{\bm{k}+\bm{\delta k}_{x}}|u_{\bm{k}}\right\rangle \left\langle u_{\bm{k}+\bm{\delta k}_{x}+\bm{\delta k}_{y}}|u_{\bm{k}+\bm{\delta k}_{x}}\right\rangle
\\ \nonumber
& \qquad \qquad \qquad
\left\langle u_{\bm{k}+\bm{\delta k}_{y}}|u_{\bm{k}+\bm{\delta k}_{x}+\bm{\delta k}_{y}}\right\rangle \left\langle u_{\bm{k}}|u_{\bm{k}+\bm{\delta k}_{y}}\right\rangle \Big).
\end{align}
A useful simplification is that the sum of the Chern numbers over a set of bands $\sum_{i=a}^{b} \mathcal{C}_i$ can be evaluated directly using an analog of \eqref{Fukui_app} in which each product $\left\langle u_{\bm{k}}|u_{\bm{k}'}\right\rangle$ is replaced by the $\det U$, where matrix $U$ is comprised of the matrix elements $U_{ij} = \left\langle u_{i, \bm{k}}|u_{j,\bm{, k}'}\right\rangle$ with $i,j = a, \dots, b$.

To facilitate the numerical procedure, it is useful to make a part of the calculation analytically. To this end, we represent each Bloch state $\left| u_{\bm{k}} \right\rangle$ as a column $(\left| w_{\bm{k}} \right\rangle, \left| v_{\bm{k}} \right\rangle)^T$ in the Nambu space. Then, one can make some analytical progress by computing the Euclidian product of the electronic ($\left| w_{\bm{k}} \right\rangle$) and hole ($\left| w_{\bm{k}} \right\rangle$) states. Recall that the cross-product $\left\langle w_{M,\bm{k}'}|v_{N,\bm{k}}\right\rangle$ does not appear since in the BdG approach, electrons and holes are independent degrees of freedom. 

Consider for the square-vortex lattice case, 
\begin{align}
& \left\langle            w_{M,\bm{k}'}|w_{N,\bm{k}}\right\rangle = \frac{1}{a_{x}}\int_{0}^{a_{x}}dx\int_{0}^{a_{y}}dy
\\ \nonumber
& \qquad \sum_{t,t'}e^{ik_{y}a_{y}t-ik_{y}'a_{y}t'}e^{-iy(k_{y}-k_{y}')}e^{ia_{y}x(t-t')/l_B^{2}}
\\ \nonumber
& \qquad \qquad \times\varphi_{N}(y-k_{x}l_B^{2}-ta_{y})\varphi_{M}(y-k_{x}'l_B^{2}-t'a_{y}). 
\end{align}
After taking the integral over $x$ and one of the sums, one can replace the remaining sum and integral over $y$ by integration over all real number with the substitution $Y = y - a_{y}t$:
\begin{align} \label{w-simplification}
    \left\langle w_{M,\bm{k}'}|w_{N,\bm{k}}\right\rangle &= \int dY\,e^{-iY(k_{y}-k_{y}')}
    \\ \nonumber
    &\qquad \quad\times\varphi_{N}(Y-k_{x}l_B^{2})\varphi_{M}(Y-k_{x}'l_B^{2}).
\end{align}
Analogously, for  holes, the expression reads:
\begin{align} \label{v-simplification}
    \left\langle v_{M,\bm{k}'}|v_{N,\bm{k}}\right\rangle  & = \int dY\,e^{-iY(k_{y}-k_{y}')}
    \\ \nonumber &
    \qquad \quad\times \varphi_{N}(Y+k_{x}l_B^{2})\varphi_{M}(Y+k_{x}'l_B^{2}).
\end{align}
These formulae can be then plugged into \eqref{Fukui_app}. The rest of the calculation is performed numerically. 

\subsection*{Sanity check for the $\Delta=0$ case }

To make sure that presented formalism is valid, we use the expressions above to calculate the Chern number of a single Landau level (in the absence of $\Delta$). The integral over $\bm{r}$ can be extended over the whole space:
\begin{align}
\mathcal{A}_{y}^{M,N} & =-i\left\langle u_{M,\bm{k}}|\partial_{k_{y}}|u_{N,\bm{k}}\right\rangle =\\ \nonumber
 & =\frac{1}{a_{x}}\int_{0}^{a_{x}}dx\int_{0}^{a_{y}}dy\,\sum_{t,t'}(a_{y}t-y) 
 \\ \nonumber
 & \qquad
 \times e^{ik_{y}a_{y}(t-t')}e^{ia_{y}x(t-t')/l_B^{2}}
 \\ \nonumber
 & \qquad
 \times\varphi_{N}(y-k_{x}l_B^{2}-ta_{y})\varphi_{M}(y-k_{x}l_B^{2}-t'a_{y}).
\end{align}
Now use the identity $\int_{0}^{a_{x}}dx\,e^{ia_{y}x(t-t')/l_B^{2}}=a_{x}\delta_{t,t'}$ and the substitution $Y = y - a_y t$ to derive
\begin{align}
\mathcal{A}_{y}^{M,N} & = \int dY\,(-Y)\varphi_{N}(Y-k_{x}l_B^{2})\varphi_{M}(Y-k_{x}l_B^{2}) 
\\ \nonumber
& = -k_{x}l_B^{2}\delta_{N,M}+\text{part, independent of }k_{x} .
\end{align}
It is also straightforward to show that $\mathcal{A}_{x}$ is independent of $k_{y}$.
So, the Berry curvature is: 
\be
\mathcal{F}_{M,M}=\partial_{k_{x}}\mathcal{A}_{y}^{M,M}-\partial_{k_{y}}\mathcal{A}_{x}^{M,M}=l_B^{2}.
\ee
Integrating this constant value over the Brillouin zone gives unity for the Chern number, as expected for a Landau level.


\section{Additional results for the triangular vortex lattice case}
\label{sec:additional-results}

\begin{figure*}
    \includegraphics[width=0.9\textwidth]{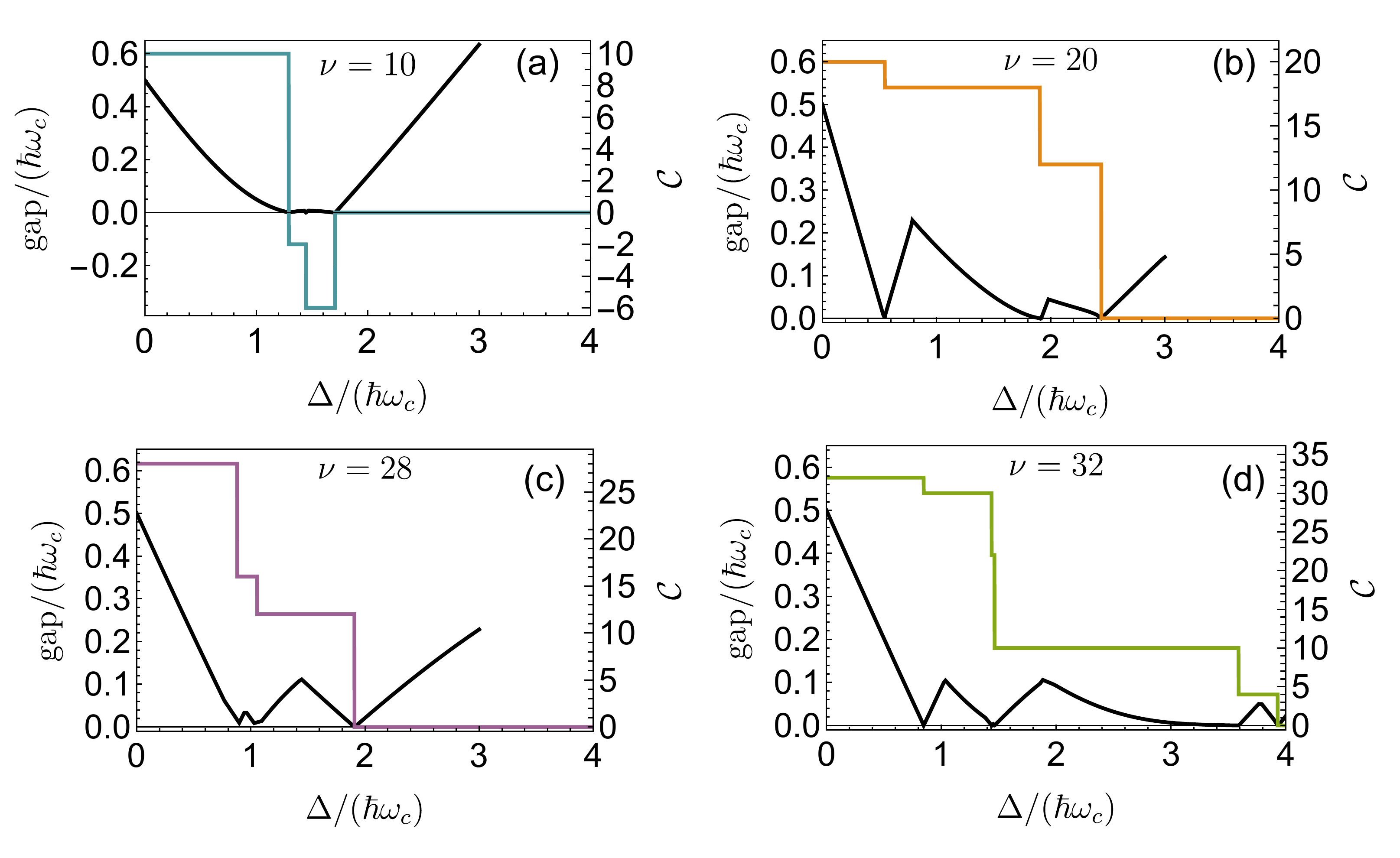}
	\caption{Quasiparticle properties in the triangular vortex lattice case for the filling factors (a) $\nu = 10$, (b) $\nu = 20$, (c) $\nu = 28$, and (d) $\nu=32$ as a function of $\Delta / (\hbar \omega_c)$, the ratio between the superconducting pairing amplitude and the cyclotron energy. Black lines, to be viewed in the left vertical scale, show the BdG quasiparticle gap. It evolves from the initial value of $0.5 \hbar \omega_c$ (midgap of Landau levels) at $\Delta = 0$, undergoes several closures at the topological transition and transforms into the de Gennes--Matricon gap at large $\Delta / (\hbar \omega_c)$. Colored lines show the evolution of the topological number, which evolves from the value of $\nu$ in the Integer Quantum Hall regime to zero in several consecutive jumps. Note that gap closures are correlated with the jumps in the topological index. 
	}
	\label{fig:triangular-additional}
\end{figure*}

In the main text, we showed in Fig.~\ref{fig:Cherns_triangular} the evolution of the topological index as a function of $\Delta / (\hbar \omega_c)$ for a number of filling factors $\nu$ in the superconducting state with a triangular vortex lattice. Here we complement these results by showing how the BdG quasiparticle gap evolves as a function of the same parameter, $\Delta / (\hbar \omega_c)$. For better visibility, we present the results for each value of $\nu$ on a separate panel in Fig.~\ref{fig:triangular-additional}.

\section{Stripe geometry calculation}
\label{app:stripe-geometry}

To study the bulk-boundary correspondence and, in particular, the existence of the edge modes, we analyze the BdG Hamiltonian \eqref{BdG} in the stripe geometry $-W/2 < y < W/2$ for the square vortex lattice case. The edges of the stripe can be modeled by a smooth potential $U(y)$ added to $H_0$, which is zero inside the stripe and grows near the boundary emulating the walls. However, since we are considering the problem in the Landau level basis, we have to compute matrix elements of $U(y)$ for wavefunctions \eqref{Landau_phi}, $U_{NM}(K_x, K_x^\prime) = \langle \phi_{N,K_x l_B^2}|U(y)|\phi_{M,-K_x^\prime l_B^2}^* \rangle $. As the wavefunctions are localized at $y \sim K_x l_B^2$, the resulting potential is going to be small at $-W/2l_B^2 \lesssim K_x \lesssim W/2l_B^2$ and growing near the edges of this segment. It is thus easier to directly introduce stripe boundaries choosing a suitable operator $U_{N,M}(K_x, K_x^\prime)$.  For instance, it can be chosen to be diagonal in the Landau level ($N,M$) and momentum $(K_x, K_x^\prime)$ spaces and modeled by the parabolic walls \cite{McDonald_new}. Using the substitution $K_x = k_x + a_y t / l_B^2$, the explicit form can be written as
\be \label{U_t_kx}
    U_t(k_x) = \begin{cases}
        0 & |\delta| < W /2, \\
        A (|\delta|-W/2)^2 & |\delta| > W /2, \\
    \end{cases}
\ee
where $\delta = k_x l_B^2 + t a_y$.

Since the boundaries violate translation symmetry in the $y$ direction, we will not use Fourier transformed wavefunctions \eqref{Landau_Fourier_psi} but rather compute the matrix elements $\langle \phi_{N,k_x l_B^2 + a_y t}|\Delta(r)|\phi_{M, k_x' l_B^2 + a_y t'}^*\rangle$ discussed in Appendix~\ref{sec:evaluating_Delta}. Using the method described therin, we get $\langle \phi_{N,k_x l_B^2 + a_y t}|\Delta(r)|\phi_{M, k_x' l_B^2 + a_y t'}^*\rangle = \delta_{k_x, k_x^\prime} \Delta_{N,t;M,t'}(k_x)$ with
\be \label{Delta_t_tp}
    \Delta_{N,t;M,t'}(k_x) = c_0 \Delta \begin{cases}
        0, & t + t' \text{ odd}, \\
        \phi_{N + M, a_y (t - t') +2 k_x  l_B^2}, & t + t' \text{ even},
    \end{cases}
\ee
with the same $c_0$ as in \eqref{Delta_NM_k} and \eqref{DeltaMN-SM}. 

With the help of Eqs.~\eqref{U_t_kx} and \eqref{Delta_t_tp}, the BdG Hamiltonian \eqref{BdG} at fixed $k_x$ becomes a matrix in spaces of  Landau levels $(N,M)$ of dimension $N_{\max} \times N_{\max}$ and in the space of $y$-coordinates of the unit cell, parametrized by integer indices $t, t'$ of the vortex lattice of dimension $W / a_y \times W / a_y$. Diagonalizing it numerically, we get the spectrum of the system as a function of $k_x \in (-\pi/a_x, \pi/a_x)$.
\begin{figure*}\includegraphics[width=0.95\textwidth]{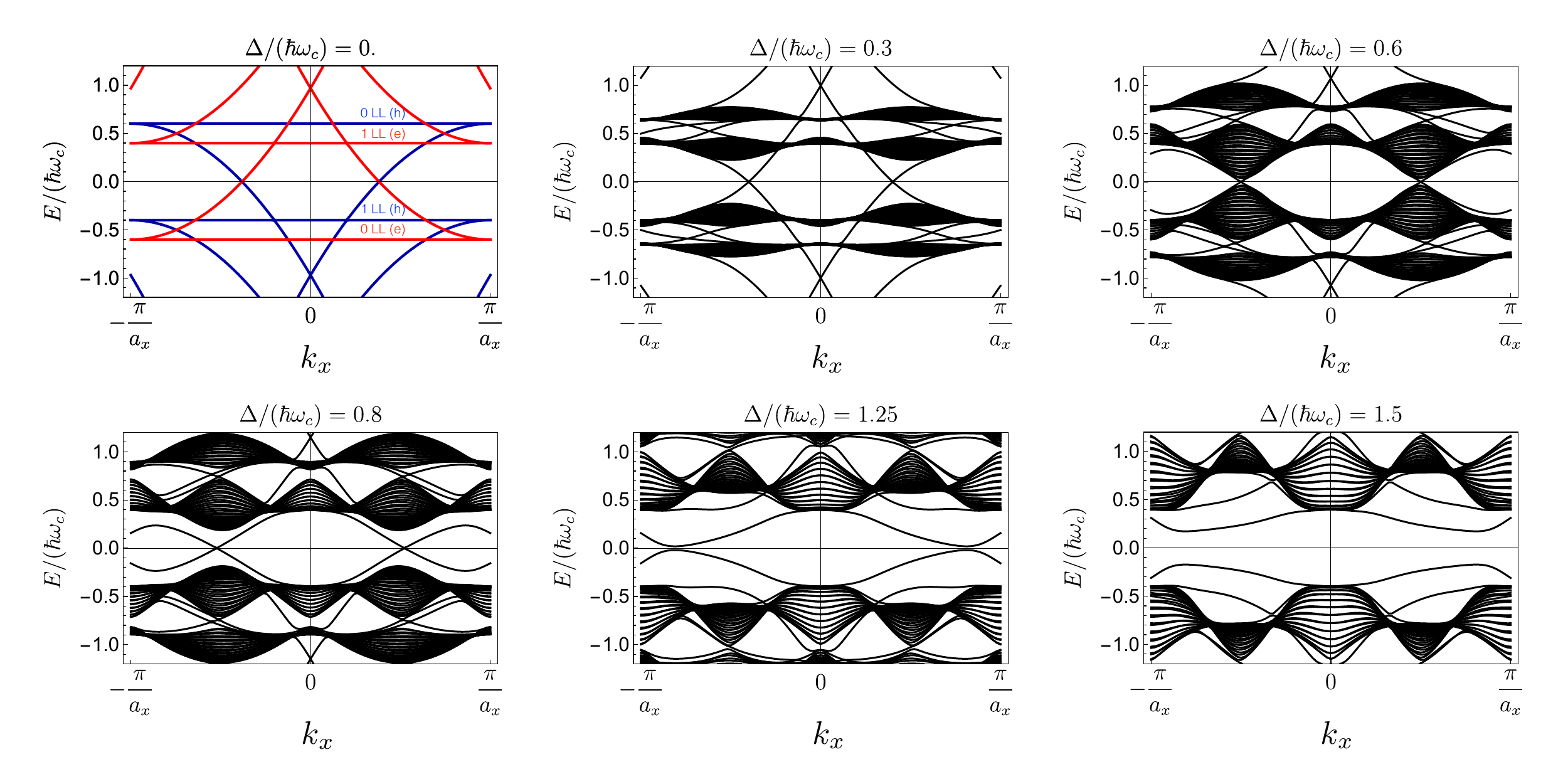}
	\caption{BdG spectrum in the stripe geometry modeled by a potential energy barriers $U(y)$ near the stripe edges, one Landau level is filled [$(E_F - \hbar \omega_c /2) / (\hbar \omega_c) = 0.6$]. The plots show the evolution of the spectrum along with the increase of $\Delta / (\hbar \omega_c)$. Quantum Hall edge modes are doubled in the BdG description. They are protected until the bulk gap closure, after which the Chern number drops to zero and the edge modes become non-topological and disappear along with the further increase of superconductivity. 
	}
	\label{fig:Stripe_9plots}
\end{figure*}

For clarity we consider a small cutoff at the second Landau level ($N_{\max} = 2$) and run the calculations at $E_F / (\hbar \omega_c) = 0.6$ filling just the lowest Landau level ($\nu=2$ if one accounts for spin). Despite small Hilbert space, this example suffices to demonstrate the edge modes evolution and agreement between the stripe and bulk calculations. In Fig.~\ref{fig:Stripe_9plots}; we show the resulting band structure for different values of $\Delta / (\hbar \omega_c$). The first panel of Fig.~\ref{fig:Stripe_9plots} clearly demonstrates flat bulk Landau level bands and the edge modes with the BdG duplication, one mode per spin per edge. At small nonzero $\Delta / (\hbar \omega_c)$, edge modes clearly survive superconducting correlations in accordance with the general topological arguments while the bulk bands acquire a nonvanishing dispersion as a function of $k_x$. The edge modes remain topologically protected until the bulk gap closure at $\Delta / (\hbar \omega_c) \approx 0.6$. This critical value matches the gap closure observed in the calculation for a bulk system. After the gap is reopened, it is no longer topological as confirmed by the bulk Chern number calculation. Accordingly, stripe geometry calculation (further panels of Fig.~\ref{fig:Stripe_9plots}) showcases that the edge modes are no longer topologically protected at $\Delta / (\hbar \omega_c) > 0.6$ and gap out with the increase of the superconducting pairing.

\bibliography{IQHE-SC}

\end{document}